\documentclass[acmsmall]{acmart}

\AtBeginDocument{%
  \providecommand\BibTeX{{%
    \normalfont B\kern-0.5em{\scshape i\kern-0.25em b}\kern-0.8em\TeX}}}

\newcommand{\major}[1]{{\textcolor{black}{#1}}}

\usepackage{CJKutf8}

\usepackage{color}

\newcommand{\hy}[1]{{\textcolor{black}{#1}}}

\AtBeginDocument{%
  \providecommand\BibTeX{{%
    \normalfont B\kern-0.5em{\scshape i\kern-0.25em b}\kern-0.8em\TeX}}}

\setcopyright{acmcopyright}
\copyrightyear{2024}
\acmYear{2024}
\acmDOI{XXXXXXX.XXXXXXX}

\begin{document}

\title[The Informal Labor of Content Creators on Xiaohongshu]{The Informal Labor of Content Creators: Situating Xiaohongshu's Key Opinion Consumers in Relationships to Marketers, Consumer Brands, and the Platform}

\author{Huiran Yi}
\email{huiran@umich.edu}
\affiliation{%
  \institution{University of Michigan}
  \city{Ann Arbor}
  \state{Michigan}
  \country{USA}
}
\author{Lu Xian}
\email{xianl@umich.edu}
\affiliation{%
  \institution{University of Michigan}
  \city{Ann Arbor}
  \state{Michigan}
  \country{USA}
}

\renewcommand{\shortauthors}{Yi and Xian}

\begin{abstract}

\end{abstract}



\received{January 2024}

\begin{abstract}
This paper critically examines flexible content creation conducted by Key Opinion Consumers (KOCs) on a prominent social media and e-commerce platform in China, Xiaohongshu (RED). Drawing on nine-month ethnographic work conducted online, we find that the production of the KOC role on RED is predicated on the interactions and negotiations among multiple stakeholders---content creators, marketers, consumer brands (corporations), and the platform. KOCs are instrumental in RED influencer marketing tactics and amplify the mundane and daily life content popular on the platform. They navigate the dynamics in the triangulated relations with other stakeholders in order to secure economic opportunities for producing advertorial content, and yet, the labor involved in producing such content is deliberately obscured to make it appear as spontaneous, ordinary user posts for the sake of marketing campaigns. Meanwhile, the commercial value of their work is often underestimated and overshadowed in corporate paperwork, platform technological mechanisms, and business models, resulting in and reinforcing inadequate recognition and compensation of KOCs. We propose the concept of ``informal labor'' to offer a new lens to understand content creation labor that is indispensable yet unrecognized by the social media industry. We advocate for a contextualized and nuanced examination of how labor is valued and compensated and urge for better protections and working conditions for informal laborers like KOCs.

\end{abstract}

\maketitle

\begin{CJK*}{UTF8}{gbsn}
\section{Introduction}

The allure of flexible content creation within the platform economy has grown significantly, especially during and after the COVID-2019 pandemic. \hy{Since} 2022, amidst a challenging job market \hy{in which} one in five of the 10.76 million college graduates \hy{in China} faced unemployment and anxiety about their future~\cite{unemployment}, the flexibility \hy{provided by platform economy} \hy{has become} particularly attractive. This flexibility, coupled with the ``flexible employment'' (灵活就业 \textit{ling huo jiu ye}), \hy{foregrounded by the Chinese government in 2023,} has shaped content production for monetization on platforms as a passion-driven \hy{work}. \hy{This is notably} distinct from traditional gig work--like that of \hy{food delivery workers, ridesharing drivers, or online pieceworkers--which is predicated on labor} outsourcing, work surveillance, \hy{body} exploitation, and the marginalization of disadvantaged classes \hy{\cite{gray2019ghost,ticona2018beyond, vallas_what_2020, jarrahi2020platformic, sannon2022privacy}}.

We focus our examination of flexible content creation on a prominent social media and e-commerce platform in China: Xiaohongshu (小红书 \textit{xiao hong shu},  which translated to ``Little Red Book,'' or RED hereafter).\hy{\footnote{This platform was established in 2013 as a community for consumers to share overseas shopping information, and it soon scaled up to an e-commerce platform in 2014. Overseas shopping, 海淘 (\textit{hai tao}), refers to the phenomenon that started to gain popularity in the 2000s when Chinese people traveled overseas and brought products back to China. Such shopping tactics help avoid taxes and maximize the value of their money~\cite{haitao}.} As a start-up that launched in 2013 and has become one of the nation's fastest-growing corporations, RED quickly caught the attention of the Chinese government and even earned praise from former Prime Minister Li Keqiang in 2015~\cite{xiaohongshu}. By 2022, RED had over 200 million monthly active users with over 43 million content creators posting on the platform~\cite{cbndata}.} Guided by its ethos of ``ordinary life's charm touches people's hearts'' (平凡烟火气，最抚凡人心 \textit{ping fan yan huo qi, zui fu fan ren xin}), \hy{RED encourages everyday users to transform their daily life experiences and practical information into monetizable content.} Following the influencer marketing practice precedents by multiple global social media platforms \cite{hund2019shoppable}, RED earns \hy{over 80\% of its} revenue through advertisement, including promotions by influencers \cite{cbndata_revenue}. RED \hy{provides} a space for users to create content with only minimal technical barriers, \hy{and it offers step-by-step guidance for those seeking more followers and progress toward influencer status.}


\hy{Creating content on RED offers a promising avenue for anxious young Chinese to monetize their creativity and earn profits by sharing their daily lives on the platform.} \hy{Yet, their efforts and the} outcomes of their labor processes are deeply entrenched in the \hy{platform's} political economy and China's broader socio-economic backdrop, especially government's policies on flexible employment. The \hy{state's goal of promoting flexible employment discourses and policies has marked a shift in the employment} status quo for young people. \hy{This change also has signaled evolving work paradigms in an economy increasingly mediated by technologies.} Our work concentrates on the often overlooked yet vital form of labor, \hy{what we call the ``informal labor'' of content creators. We focus on a specific group of content creators on RED---key opinion consumers (KOCs)---who typically have between 5,000 and 20,000 followers and are important for marketing campaigns on RED. Although these KOCs appear to be everyday users, they yet get involved in posting product reviews and recommendations to other platform users for advertorial purposes.} We articulate how such content creation labor has remained informal and largely unseen \hy{in RED's influencer marketing campaigns}.

\hy{Drawing on feminist sensibilities in human-computer interaction \cite{bardzell_feminist_2010}, we attend to the digital platforms and cultural work with an alternative standpoint that privileges situated knowledge\cite{tsing2015mushroom, haraway2013situated}. Thus, we pay attention to the less noticeable and often underrepresented subjects and content creators' experiences. At the center of our investigation are two main research questions: \textit{How do platform users take upon the particular role of KOC? What kinds of trivial and mundane labor of content creators is required for digital cultural production on platforms?}} \hy{By delving into the dynamics among content creators, marketing practitioners (marketers), and digital platforms, we highlight the fact that despite the challenges KOCs face when creating monetizable content, this type of informal labor empowers young people in China. This is especially significant, considering the current high unemployment rate and precarious job market.}

With our analysis of KOC labor in the platform economy on RED in China, we make three primary contributions to the computer-supported cooperative work (CSCW) community. \major{First, by examining labor processes and the ways laborers are disadvantaged, we engage with the ongoing discussion in CSCW about the need to revisit the concept of labor \cite{greenbaum_back_1996, tang_back_2023}. Specifically, we showcase how the political, economic, and platform-mediated relations orchestrate the activities of various stakeholders as exemplified by KOC content creation.
Second, we articulate how such labor---which involves both aspirational \cite{duffy_romance_2016} and tactical efforts---is informalized and devalued in platform-mediated marketing campaigns, platforms' technological mechanisms, and business models. 
The ``informal labor'' concept we propose offers a new lens to understand content creation labor that is indispensable yet unrecognized by the social media industry. 
Third, our research explores the implications of such informal labor on individual lives in light of China’s ``flexible employment'' policies and discourses, as well as young Chinese people’s anxiety about post-pandemic job insecurity. 
We show that such informal labor, to a certain extent, equips young people with new tools and skills that help them navigate the employment crisis.} 


\major{\section{Background}

\subsection{China's ``Flexible Employment'' Policies in the Time of an ``Internet Plus'' Economy}


We situate this research against the backdrop of \major{the mounting unemployment among young, well-educated Chinese college graduates, which has escalated since the COVID-19 pandemic began in 2019. 
The recent rise of generative AI, 
widespread lay-offs, and global economic downturn have intensified this crisis ~\cite{zhang2023, BigTechcrackdown, layoff}. The national unemployment rate among urban young people aged 19 to 24 soared to a record-breaking 21.3 percent in June 2023~\cite{fu2023}. In response to the accompanying mass unemployment anxiety, the Chinese government emphasized ``flexible employment'' (灵活就业\textit{ling huo jiu ye}) as a key strategy for economic rebound and social stability, and they promoted the subject through state mouthpiece media \cite{flexibleemployment}.}

While it has evolved since the onset of the pandemic, China’s initial discourse on flexible employment, and its related policies, emerged in the late 1990s and early 2000s in order to pacify large-scale job losses that resulted from economic restructuring and privatization.
By 2003, more than 28 million workers, particularly former state employees previously guaranteed lifetime employment with their ``iron rice bowl'' jobs (铁饭碗 \textit{tie fan wan}),
became unemployed~\cite{zhong2004, xiagang}. 
Amid their growing unrest and widespread despair \cite{yang_unknotting_2015}, the government initiated the 
``reemployment'' (下岗再就业 \textit{xia gang zai jiu ye}) project~\cite{solinger_creation_2006, yang_unknotting_2015} to support its citizens, and redefined employment to include ``part-time, temporary, seasonal, and flexible work'' through the ``flexible employment'' discourse and related policies enacted in 2002 \cite{shi2011}. Those kinds of work arrangements---flexible, piece-wise, and short-term---departed from the traditional, rigid schedules of state employees. 
The government also bolstered these efforts by updating social security (社会保障 \textit{she hui bao zhang}) measures to guarantee the welfare of people in flexible employment \cite{li2002,shi2011} in the early 2000s. Since then, new forms of work--such as piecework \cite{irani2015difference}, gig work \cite{gray2019ghost}, and entrepreneurship \cite{avle2016design}--have become prevalent, particularly among young working-class individuals, including migrant workers and those with lower educational \hy{degrees}. At the end of 2021, approximately 200 million Chinese 
were identified as flexibly employed laborers \cite{zhong2021}. 

Entering the 2010s, the Internet technologies and platform economies further updated flexible employment opportunities. Since 2015, the State Council has championed the ``Internet Plus Economy'' (互联网+经济\textit{hu lian wang jia jing ji}) initiative, which leverages Internet technologies to stimulate economic growth across various sectors, which includes the development of social commerce(社交电商\textit{she jiao dian shang}) \cite{guang2015, zhong2015} via social media networks. In 2021, the Chinese Ministry of Human Resources and Social Security officially recognized ``Internet Marketing Specialist'' (互联网营销师 \textit{hu lian wang ying xiao shi}) as a profession in its \textit{National Occupational Classification} (国家职业分类大典\textit{guo jia zhi ye fen lei da dian}) \cite{zhong2021-2} released by the Chinese Ministry of Human Resources and Social Security. This category encompasses roles such as 
e-commerce live-streamers, product selectors, platform administrators, digital content creators, and promoters \cite{zhong2021-2}. The state’s formal recognition identified this group of laborers as instrumental in boosting the interactivity and credibility of online platforms, especially related to marketing and promoting corporate products. This fact underscores the important role that Internet marketing specialists have played in the recent stimulation of domestic consumption of goods. 
State-owned media, including \textit{Xinhua News}, have highlighted success stories of live streamers, encouraging young people to engage with and propel the e-commerce and platform economy \cite{jing2023}.

Meanwhile, the unemployment crisis among college graduates persists. In response, other state media outlets like \textit{the People's Daily} have urged young Chinese to pursue ``individualized (个性化\textit{ge xing hua}) jobs or careers out of their autonomy (自主性\textit{zi zhu xing})'' and have noted that sectors such as platforms, the cultural and creative industries, service sectors, and the knowledge economy are ripe for innovation and employment \cite{li2022}. 
This narrative steers the young Chinese away from traditional, ``iron-rice-bowl'' jobs towards more flexible careers. 
Occupations like Internet marketing specialists have surfaced as ideal roles within this new paradigm, with popular platforms like Xiaohongshu (RED), Douyin, Kuaishou, Bilibili, and WeChat as desired spaces to start flexible careers.

\subsection{Xiaohongshu (RED), Influencer Marketing, and Key Opinion Consumers (KOC)}

We focus on Xiaohongshu (RED), a key player in China's rising platform economy. RED started its business with the dual goal of creating an online community and an e-commerce space, as co-founder Qu Fang \cite{xiaohongshushinian} believed that consumers are likely to be influenced by people they know and trust, and well-informed consumers make better purchasing decisions. 
As RED scaled up, a particular marketing discourse and product recommendation practices, ``The Art of Planting A Grass'' (种草学\textit{zhong cao xue}) emerged. This discourse emphasized the platform's goal of promoting commercialization that hinges on influencer marketing practices (商业化\textit{shang ye hua cao xue}) while maintaining its``community-based content ecology'' (社区内容生态\textit{zhong cao xue})\cite{xiaohongshushinian}.
In 2023, RED launched a learning platform under the name ``the art of planting a grass,'' which was designed to enhance platform users’ community-based marketing skills by education them about a broad range of topics from photo and video editing to creating viral content and monetizing online presence and was designed to enhance platform users' community-based marketing skills \cite{xiaohongzhongcaoxue}.
We consider RED's ``art of planting grass'' concept to be parallel to ``influencer marketing'' more often seen in the Western context. Influencer marketing refers to sponsoring social media users and leveraging their performed identity to promote brands, as seen in Instagram \cite{hund2019shoppable, influencermarketing, hund2019shoppable, forbes}.
We use influencer marketing later in this paper to describe the interactions between marketers and content creators on RED.

Key Opinion Consumer (KOC) was first invented by RED for influencer marketing purposes particularly, which distinguishes itself from the more recognized concept of key opinion leader (KOL), 
which refers to influencers in the Western context~\cite{hund2019shoppable, kol} who typically have many followers (usually at least 20,000) and monetize their content on platforms \cite{abidin2016aren}; KOLs are often viewed as crafted personas for mass consumption
\cite{turner_approaching_2010}. 
In contrast, KOCs are seen as 
ordinary, unembellished, and relatable individuals 
with followers ranging from 5,000 to 20,000 in the case of the RED platform. 
For marketers who work for corporations that invest in influencer marketing campaigns, KOCs are friend-like amateurs and ``simple, unpolished people'' 
(素人\textit{su ren}) whose experiences are likely to impact consumers’ purchasing decisions \cite{smith2019}. 
Chinese cosmetic brand Perfect Diary pioneered the leveraging of KOCs for marketing on RED, demonstrating that KOCs could effectively disseminate marketing messages and key selling points to potential customers. The strategy to involce KOCs in marketing campaigns quickly spread across China's platform economy to other platforms, such as Douyin (Chinese TikTok and Taobao). Even though marketers see KOCs as \textit{su ren}, they are distinct from ordinary platform users because of their ability to monetize content on the platform by collaborating with consumer brands.



}
\major{\section{Related Work}


\subsection{The Focus on Labor Processes in Platform-mediated Work}

\major{Greenbaum \cite{greenbaum_back_1996} called on the CSCW community in 1996 to attend to not only the skills and work practices but ``wages, working conditions, and labor relations'' (p.234). Decades later, in 2023, CSCW scholars \cite{tang_back_2023} returned to this topic to raise researchers’ and the public’s awareness of the political and economic ramifications of work. This focus on labor processes underscores the importance of uncovering the dynamics within workspaces and unraveling the division and coordination of labor that underpins the performed work \cite{greenbaum_back_1996}.}

Shifting the focus from work to labor illuminates the underlying social and labor relations.
Researchers within the CSCW community and related fields have advocated for more humanized work conditions and improved experiences for food delivery workers \cite{chen2022mixed} and inclusive and collective solidarity among on-demand platform workers \cite{yao2021together}; they have also examined the new workplace management tools and practices \cite{khovanskaya2019tools} and designed systems that support laborers by intervening labor market and relations \cite{dombrowski2017low, gloss2016designing, spektor2023designing, freeman2020mitigating}. \major{In particular, centering laborers' experiences brings to light ways to improve their working conditions. In the case of Amazon Mechanical Turk workers, scholarly work that uncovers their experiences and relationships with task requesters \cite{martin2014being, salehi2014dynamo} and designs systems to support them rate task requesters \cite{irani2013turkopticon, irani2016stories} contributes to greater recognition of their work and addresses issues of invisibility.}

\major{Focusing on labor processes also allows for an understanding of the nuanced and often invisible forms of labor within platform-mediated work. Gig and platform workers often remain unseen and unrecognized by end users, a design feature of the platforms themselves \cite{vallas_what_2020, barratt_im_2020}. The concept of invisible labor, characterized by its economic undervaluation \cite{hatton2017mechanisms}, includes activities performed within paid employment in response to employer's both implicit and explicit demands \cite{crain2016invisible}. Focusing on crowd gig work like Turkers' piecewise work specifically, Gray and Suri \cite{gray2016crowd} show that such unpaid labor, which is increasingly driven by technological advancements, often resulted from the tasks that companies traditionally managed for workers, such as searching for desirable tasks, reading task instructions, and managing work breaks \cite{gray2016crowd, sannon2019privacy, gadiraju2017modus}. To highlight issues of fair compensation, Toxtli et al. \cite{toxtli2021quantifying} quantified the unpaid, invisible labor of crowd workers.
In addition to the forms of invisible labor inherent in the processes of performing labor, CSCW scholars have underscored the often overlooked emotional labor in digital workspaces and platform capitalism. For example, Raval and Dourish \cite{raval2016standing} demonstrated through ethnographic research how ridesharing platforms' algorithms strategically manage gig workers' emotions and physical presence, yet this emotional labor is rarely acknowledged within formal job descriptions and thus undervalued \cite{star1999layers}.}

\major{In line with CSCW's long interest in labor processes and invisible labor, we attend to a particular form of invisible labor on social media platforms---that of content creators. Different from the focus on the tasks and workplace arrangements that render platform-mediated labor invisible, we explore how the concrete working practices of corporations and platforms informalize content creation labor and render it invisible in the case of key opinion consumers on RED.}

\subsection{Labor on Social Media Platforms: Visibility and Authenticity}


\major{Existing literature in communication and media studies on gig work within social media platforms often examines digital content production as cultural commodities \cite{nieborg_platformization_2018}.
Social media influencers contribute essentially to digital content on platforms, and they are social media users who blend their daily posting with marketing communications to produce advertorial content \cite{abidin2016visibility}. As entrepreneurial laborers, they live on their creativity while undertaking piecemeal tasks \cite{neff2005entrepreneurial}. Duffy \cite{duffy_romance_2016} highlights the aspirational aspect of influencer labor through a gendered lens, suggesting that while this work offers women a departure from patriarchal employment structures, it remains largely unrecognized, uncompensated and perpetuates gendered social hierarchies.
Abidin argues that platform algorithms transform influencer activities into 
``visibility labor'' \cite{abidin2016visibility}, as the platform economy is predicated on sustained attention. To maintain their visibility and engage their audience, influencers must continuously adapt to technological advances while creatively integrating commercials into their content creation \cite{abidin2016visibility, nieborg_platformization_2018}.} 

\major{We draw on literature in communication and media studies and CSCW/HCI that underscores the importance of visibility management. 
Algorithms, as opaque bureaucratic systems \cite{seaver2017algorithms, ma2022not}, shape content curation and creator management.
As a result, influencers have to research and devise strategies to enhance and sustain their visibility. 
Concepts like ``algorithms gossip'' discussed by Bishop \cite{bishop2019managing} and ``algorithmic imaginary'' introduced by Bucher \cite{bucher2019algorithmic} highlight how influencers collectively share resources to understand algorithms and their changes. Ordinary users also grapple with algorithms that affect visibility, often experiencing platform moderation as adversarial since it can render them invisible \cite{ma2023users}. 
Duffy and colleagues \cite{duffy2023platform} point out that social factors are critical in enabling or thwarting users' visibility, and often systematically discriminate against marginalized identities.}

\major{Within the social media influencer culture, influencers' value is tied to their performed authenticity. Scholars in popular cultural studies argue that authenticity is often performed for economic values and exchanges \cite{banet2012authentic}. For influencers, authenticity is also predicated on the performances of amateurism and relatability \cite{duffy2017not}, and the ``authenticity ideal'' specifically emphasizes ordinariness \cite{hund2019shoppable} with limited expression of expertise and professionalism \cite{craig2019social}. Influencers who have to appear relatable and authentic to everyday platform users and at the same time incorporate advertisement in their content, stand in a distorted position. They have to be commercially valuable but conduct a performance that comes from their true self \cite{bishop2023influencer}. Authenticity, for influencers, is desirable yet risky. For them, this authenticity is commercially advantageous for opportunities to participate in marketing campaigns, yet personally risky in that the performances of authenticity impact their personal lives and emotions \cite{bishop2023influencer}.}

\major{Our work enriches the existing scholarly discussion on visibility and authenticity in influencer culture by identifying and discussing mechanisms that govern digital content creation in the case of RED. 
We investigate the emerging role of KOCs in content creation for monetization and examine 
the relations between their true selves as an everyday person and what they perform in their content on the platform.}

}
\section{Methods}

\subsection{Data Collection}

This paper draws on long-term digital ethnographic research that the first author conducted on the RED platform, including digital fieldwork between December 2021 and September 2023. \major{Digital ethnographic research aims to understand ``shared practices, meanings, and social contexts, and the interrelations among them'' of phenomena predominately taking place through digital means \cite[p.67]{taylor2013ethnography}. The digital fieldwork focuses on two main items: 1) the practices of individuals who create content and pursue monetization through RED's influencer marketing campaigns, and 2) the practices of other actors who facilitate these marketing efforts. 
Additionally, our analysis incorporates contextual data about RED, including the company’s mission statement, the Chinese government's policies on digital platforms and the coverage in both Chinese and international media, and public statements by RED’s management teams. This broader context helps us understand how the practices and sense-making of content creation by various actors are framed within the prevailing discourse of flexible employment. The study was granted an exemption by our university's institutional review board.}

\major{The empirical data for this paper was derived from the first author's fieldwork, during which they engaged in participant observation on the RED platform. This involved creating an account to like, collect, and comment on advertorial posts, as well as interacting with other users through comment replies and private messages. 
Through these interactions, the first author gained substantial insights into advertorial content on RED, including its formats (e.g., video, photos with text), topics, and user reactions.
Additionally, the first author leveraged this platform to connect with content creators and establish deeper interactions through private messages on RED as well as messaging tools like WeChat and iMessage, and in-person meetings. This groundwork supported the conducting of interviews with 16 RED content creators. }
Due to travel limitations caused by COVID-19, some of the interviews happened via Zoom. 
\major{Building on her established networks from previous roles at multinational corporations and through content creator informants, the first author also connected with key industry figures involved in} RED's influencer marketing campaigns 
\major{including} marketers, advertising agency associates, and employees of the RED platform company \major{in Shanghai, resulting in interviews with eight such informants}. 

\major{The interview questions were tailored to the roles of the interviewees. 
Content creators were asked about their motivations, experiences creating advertorials for commercial brands, and the challenges they face, such as brand collaboration, navigating RED's technological design, and understanding and aligning with RED's cultural values. The questions for other informants like industry specialists focused on their strategies for selecting and interacting with content creators and the platform.} 
At the same time, the first author \major{looked into the personal accounts of the 16 content creator informants and examined their posts created for various brands for advertorial purposes.} \major{Through the research, any personal identifying information such as names, job titles, and corporation names are anonymized or removed.}

\major{\subsection{Data Analysis}}

The first author took fieldnotes and wrote memos throughout her fieldwork. \major{Fieldnotes record researchers' understanding of the particular moment in a given field site, and they can be both the descriptions of the site and initial interpretations of it \cite{salmons2016collecting}. Sequential to fieldnotes, the first author wrote memos. Memos help to further interpret and analyze the research, for instance, memos unpack personal reflections and elaborate on collected data and modes of analysis \cite{emerson2009shaw, mohajan2022memo}. During the memo-writing process, the first author reflected on the particular terms and phrases that helped to elaborate on the theme.} 

Both authors worked together to code the data and wrote ethnographic vignettes in addition to weekly meetings and discussions. \major{We coded the data by using grounded theory techniques to capture the complexities in interview transcripts~\cite{clarke2015situational}, and the analysis was conducted over 8 months from May 2023 to December 2023.} \major{We first analyzed the sample interview independently through open coding to familiarize us with the data and concepts. In a similar fashion, we each coded the sample fieldnotes. After this round of coding, we met and discussed our codes and why they were appropriate.} In this round of coding, the following key descriptions emerged: life experience as practical information; content monetization; marketing strategies; platform’s cultural values and curating algorithms; and labor in organization. \major{We made a list of codes that we both agreed upon. Then, we did a second round of coding with the rest interviews and fieldnotes. We discussed how to use codes, investigated the conflicts, and made a codebook. In this codebook, we have some main themes emerged,} including ``Key Opinion Consumers,'' ``practical information,'' ``monetization,'' ``vertical,'' ``side hustle,''``marketing tactic,'' ``corporation procedures'' (in parallel with ``corporation paperwork/clerical work''), ``labor,'' and so on. \major{These themes help us structure the Findings. For example, ``vertical'' went to \ref{sec:cater-needs} and ``corporation procedures'' became \ref{sec:corporation-measurement}.}

These themes prompted both authors to consider the political-economic context in which the labor of content creation was situated. \major{We conducted discourse analysis \cite{kackman2018craft, gill2000discourse}.} By examining online archival documents regarding China's platform economy and the state government’s policies on employment and Internet Plus and \major{mouthpiece} media news reports, \major{we learned how historical state policies that promoted flexible employment and the platform enacted flexible work. We discerned that the content creators were part of the large-scale phenomenon of flexible employment particularly in the time during and after COVID-19, amid global economic hardships and rising unemployment rates.} \major{Contextualizing the content creators' and other actors' practices in the backdrop of the Chinese state's discourse and policies on flexible employment, we are able to understand the political, economic, and social meaning of these practices and theorize how and why content creators' labor is important.}


\subsection{Positionality}

The research was conducted by two authors, with the first having professional experience as a marketer for multinational corporations prior to her PhD studies. Her experience as a social media marketing specialist positions her as an insider--a member of the content creation community. She is well aware of the practices of multiple actors put into making KOCs. \major{Furthermore, she has witnessed content creators, especially KOCs---despite the time and effort they invest in performing persons and seeking monetization opportunities---have been rendered invisible in social media influencer marketing.} The first author also noticed that many content creators who competed for the KOC roles took it as a career starter after graduation or as a side job supplementing their primary full-time employment. This complexity motivated her to investigate the practices of the content creators, asking how they arrived at the KOC stage. The first author follows the feminist ethnography sense of reciprocity \cite{leavy2018contemporary} to create critical distance so as to observe the phenomenon as an outsider. 

The second author is an active user of the RED platform and has leveraged its convenience, utility, and practical knowledge in the content created by both KOCs and users. In addition, she has noted the growing trend among young people, who are grappling with post-pandemic employment challenges, to aspire to become KOCs and other semi-celebrities on online platforms. This observation has spurred her to investigate further into the monetization of online presence, the dynamics at play between KOCs and the platform, and the complexities involved in the content creation process. 

\major{\section{Findings}
We illustrate the processes through which content creators seek economic opportunities in China's platform-mediated advertorial content creation market, often by acting as KOCs. 
The opaque and idiosyncratic paths toward monetizing their digital content creation and transforming themselves from everyday users/creators into KOCs require them to make sense of the digital platform's technological design and business models, accumulate a substantial number of followers, and sustain their visibility. 

The production of the KOC role on the RED platform is predicated on the interactions and negotiations among multiple stakeholders. The co-production of KOCs is situated at the intersection of several key actions: content creators' adoption and performance of platform-curated personas (Section~\ref{sec:koc-personas}) and their strategy of catering to followers' interests (Section~\ref{sec:cater-needs}), the concrete practices of marketers who select content creators in accordance with their business needs (Section~\ref{sec:marketing-practice}), the budget-allocating metrics used by corporations (Section~\ref{sec:corporation-measurement}), and the RED platform's content governance tools (Section~\ref{sec:red-governance}). 
We highlight that the content creators are often subordinated to the management of marketers, corporations, and platforms. Despite this uneven power distribution among multiple stakeholders, becoming a KOC on RED is, to some extent, empowering for these content creators in the concurrent political-economic context of joblessness and anxiety among young Chinese (Section~\ref{sec:(dis)empowerment}).

\subsection{Becoming a Key Opinion Consumer}

\subsubsection{Performing Platform Curated Personas}\label{sec:koc-personas}

Adopting and performing a persona to show authenticity is not a novel practice in the digital content creation industry. In our research, we have observed that personas are not necessarily distinct from their daily life experience.
For content creators who start from scratch, the way they gained popularity impacted what kinds of persona they later chose to show on the platform.
The platform's algorithm played a critical role in curating the personas and interpellating them into content creators' ideas and practices, and creators perform these personas to align with the platform’s expectations.  

Lexie started her RED journey as a creator on RED by sharing her Outfit Of The Day (OOTD) as a young professional woman in her early 20s. It was a post about her job interview at a Global Fortune 500 multinational corporation that unexpectedly went viral overnight. For Lexie, gaining followers happened unexpectedly and outside of her control. She recalled, \textit{``I spontaneously posted on RED because I was invited by the human resources to record a campus recruitment video, and that posting suddenly went virtual.''} 
She quickly learned that the corporation's name is ``visibility secret'' --- it was favored by the platform's algorithms so that her posts could reach a wide audience's content feeds on RED.
She then started posting about her workplace experiences at the multinational corporation and sharing her thoughts, and reflections, embracing the platform-curated persona of a ``workplace content creator'' (职场博主\textit{zhi chang bo zhu}). 

From Lexie’s perspective, unexpectedly becoming a workplace content creator due to the platform's obscure algorithm was a stroke of good luck, as she did not intend to pursue this role initially. Performing this persona brought her feelings of self-actualization. For instance, she noted, \textit{``When my followers commented that they found my posting helped them to deal with job interviews or improve work efficiency, I felt that all my efforts were meaningful.''} Lexie found her niche genuinely impactful, in stark contrast to the stylish but vapid Instagram content creators, with whom she had previously been fascinated. She found their beautiful daily outfit styles repetitive and unhelpful, and she could never become as beautiful as they appeared. Lexie explained, sharing practical workplace advice helped her followers, particularly young professionals like herself, to facilitate their real professional and personal growth and improvement.

When asked about her ultimate reason for posting on RED, Lexie did not hesitate, ``Making money! (搞钱~\textit{gao qian}).'' Success on RED, like on any social media platform, hinges heavily on follower count. More followers increase potential visibility, which is supported or thwarted by the platform algorithms. Working in the advertising industry, Lexie knew how to effectively engage her audience by using influencer marketing strategies. Aiming to ``make money,'' Lexie focused on quickly increasing her follower base, but she also identified another key tactic in addition to the performance of the platform-curated persona: performing ``beauty.'' She used ``good-looking'' portrait photos as cover images for her posts to draw the audience's attention and encourage clicks. In her view, ``good-looking'' meant appearing healthy and energetic, which contrasted sharply with the prevailing Chinese internet aesthetic that favored fair skin, youthful face, and slim figure \hy{\footnote{Fair skin, youthful face, and slim figure is a direct translation of 白幼瘦 (\textit{bai you shou}), a popular Internet discourse in China that describes the appearance of women. The preference for this aesthetic is sometimes critiqued as unhealthy}.} Lexie confidently affirmed her perspective, \textit{``Toned and tight muscular is more attractive.''} In her cover photos, Lexie showcased her exercise-toned arms and shoulders, complementing these with a radiant, big smile. These all reinforced her vibrant, smart, and energetic workplace persona, matching her professional image as a workplace content creator.

Lexie is not alone in performing a platform-curated persona. Shirley, who lives in Dallas, USA, and frequently travels back to her hometown, Shanghai, China, similarly stumbled into the world of content creation on RED. Like Lexie, Shirley did not expect to gain hundreds of ``likes'' on the platform overnight; rather, she began using RED to kill time when working from home during the COVID-19 pandemic in late 2020. Her posts about Japanese restaurants in Dallas garnered RED users' attention, which motivated her to create more videos showcasing local restaurants in Dallas. Shirley did not think of herself as a heavy social media user before discovering RED. She found traditional social media platforms (such as WeChat and Weibo, two prominent social media platforms in China) with their often inauthentic content and unrealistic beauty standards to be sources of anxiety. RED struck a different chord with Shirley. She explains, \textit{``I found people on RED are more likely to `be oneself' and `not compare with others.''} The platform's emphasis on mundane, everyday content appeals to users like Shirley, who generally shun the flashier side of social media. Shirley made content based on utility and relevance for her followers in order to ensure that her posts were genuinely helpful. She noted, \textit{``I believe that only `practical information (干货\textit{gan huo})' could be beneficial to my followers and keep them with me.''} In performing the persona of a local guide, Shirley tailored her posts to offer what she considered truly valuable information for Chinese residents and tourists in Dallas.

Content creators like Lexie and Shirley, who inadvertently found themselves in the realm of content creation, took advantage of the surge in their popularity by adopting the personas curated by the platform to sustain their visibility. For them, performing personas required them to truly think in a way that they and their personas were connected. It is a continuous effort to understand what does or does not resonate with their followers, especially when navigating the platform's opaque visibility algorithms with the goal of monetization on the platform.

\subsubsection{Toward Monetization: Catering to Followers' Interests and Producing Niche Content}\label{sec:cater-needs}

On the path to becoming a KOC who can monetize on RED, a creator needs to produce themed content in alignment with their personas. We will illustrate below how creators skillfully navigate the platform so as to identify the themes of their content creation, position themselves as marketable entities, and seek monetization opportunities by collaborating with potential advertisers.

Dedicated to performing her workplace professional persona, Lexie leveraged the demographic features of her followers--analytics that RED offers to creators--to tailor her content. These analytics showed that half of her followers were aged between 18 and 24, 35\% lived in Shanghai, and their primary interest was in ``education and knowledge.'' Recognizing that many of her followers were young professionals and recent graduates entering the job market, Lexie honed her content strategy around three key hashtags: ``fast-moving consumer goods industry knowledge,'' ``Fall campus recruitment interview Bible,'' and ``marketing in daily life,'' with each targeting a specific segment: industry insights for young professionals, interview tips for recent graduates, and marketing trivia for a general audience, respectively. Lexie prioritized making her content ``tangible:'' \textit{``Being tangible is being concrete [...] I hope I am tangible, like a friend in my followers' life, not someone aloof.''}  This meant providing detailed, mundane posts relevant to ordinary people. Lexie believed that this concreteness would cement her as a (seemingly) credible expert on topics that her followers were interested in. As she explained, this also reflected her philosophy of ``growing with my followers,'' which she achieved by producing work-related niche content that strengthened her persona and garnered sustained attention from her followers, a tactic known as ``being vertical'' (垂直\textit{chui zhi}); she aligned content with her curated persona and her follower interests \cite{lobato2016cultural, bishop2021influencer, hund2019shoppable}. 

Such verticality attracted the attention of advertisers. After reaching the 10,000-follower threshold, Lexie was contacted by talent brokers, often known as multi-channel network (MCN) agencies in China. One of the agencies connected Lexie with a computer monitor brand. The computer monitor was a natural fit for her videos, which frequently showed her working with her computer in the background. The promotional videos about her work experience videos blended seamlessly into a single post.

Jasper is a young man in his late 20s, working at a cosmetic company in Shanghai, and also a content creator on RED. He agreed that ``being vertical'' could lead to more monetization opportunities on RED. He upheld his persona as a ``good-looking'' (高颜值\textit{gao yan zhi}) content creator, sharing stylish daily outfits and lifestyle photos. To solidify his good looks, his account was neatly organized, featuring a consistent aesthetic with simple, film-like filters and images of himself in tastefully decorated indoor environments. For Jasper, a clean and fit look not only kept his followers engaged but also reminded them of ``a pleasant scent.'' Being vertical, in Jasper's words, \textit{``increase(s) my credibility.''} Such credibility is gained through claiming expertise and specializing in content related to personal aesthetics. This expertise reinforced his persona, creating a self-sustaining cycle of credibility and attractiveness. Jasper emphasized the importance of being vertical: \textit{``Vertical account is more expensive.''} 
By ``expensive,'' Jasper meant both its practical utility and market value. Brands prefer to partner with vertical accounts because they are seen by marketers as more likely to influence consumer behavior on RED, and vertical content creators often receive greater financial rewards.

\subsection{The Co-production of Key Opinion Consumers: A Marketing Tactic}

To become a KOC on RED, creators adopt and perform platform-curated personas, cater to their followers' interests, and produce focused niche content. However, the effort involved in preparing such content does not guarantee success, as the production of a KOC is a collaborative process involving multiple actors. We argue that in order to understand the making of the KOC role, we must situate KOCs in the context of their relationships with marketers who work for consumer product brands, corporations that invest in marketing campaigns, and platform governance.

\subsubsection{The Marketer's Expectations for Content Creators}\label{sec:marketing-practice}

Marketers usually collaborate with a large number of content creators at large-scale marketing campaigns. Without participating in these campaigns, content creators cannot monetize. While marketers' practices are critical for KOCs, the specific criteria for KOC selection fluctuate. 

According to Celia, a social media marketing specialist at a multinational cosmetics company in Shanghai, using KOCs in marketing campaigns is a cost-effective strategy with high potential returns. Rather than seeking high-quality advertorial content from KOCs, marketers focus more on the number of KOCs involved in each campaign. This strategy, described by Celia as ``casting a wide net'' (广撒网\textit{guang sa wang}), involves investing in many KOCs who create themed advertorial content during a campaign to maximize coverage across users’ customized RED feeds. For marketers, the more KOCs who are involved, the more they can ``preheat'' (预热\textit{yu re}) the platform, familiarizing potential consumers with the products and boosting the volume of online discussion before the grand campaign launch.
Additionally, collaborating with KOCs generates earned media exposure in terms of RED users'  likes, reposts, shares, comments, and other interactions. Marketers value this type of exposure, likening it to ``tap water'' (自来水\textit{zi lai shui}), inexpensive or even free, yet essential. The seemingly unembellished content produced by KOCs is expected to be relatable to a wide audience. Interestingly, the term ``\textit{zi lai}'' also implies automation in Chinese. This analogy highlights marketers' expectations for KOCs to be self-sustaining and low-maintenance, appearing effortlessly automated.

The KOC role is fundamentally relational, only established upon a content creator's position within a marketing campaign. Such a position is not only predicated on the creators' own preparation as discussed earlier, but also on decisions made by marketers. For marketers, the process of identifying a sufficient number of suitable KOCs involves tedious, time-consuming labor. They need to ensure there are enough qualified content creators who can serve as KOCs for their campaigns. Additionally, they need to align these KOCs with the brand's messaging goals to potential consumers on particular social media platforms in order to effectively influence consumer purchasing decisions. 

Advertisement agencies specializing in social media marketing, hereafter referred to as ``social agencies,'' often provide marketers with lists of potential content creators to marketers based on their specifications. Social agencies identify such creators and source potentially suitable ones, who they consider for KOC roles, by ranking their follower numbers available through third-party data providers and by collaborating with MCN agencies that have direct contact with content creators.\footnote{Content creators ranking by their follower numbers can be found through RED's third-party data providers, such as \hyperlink{qian-gua.com}{qian-gua.com}.} 

Once a broad pool of content creators is assembled, the marketers then determine which creators qualify for the campaign. They typically go through pages of spreadsheets and investigate candidates' account profiles as well as content, with the goal of building the brand's own ``KOC talent pool'' in mind. When asked about the criteria for a KOC, Celia answered without hesitation, \textit{``Someone who appears attractive and offers persuasive perspectives.''} Attractiveness is subjective and largely depends on the marketer's personal judgment, making the selection criteria volatile and reflecting biases and discrimination towards one's appearance prevalent on social media platforms \cite{duffy_having_2015}. 

Persuasiveness is another criterion, which pertains to whether a KOC's content is ``vertical.'' Celia elaborated, stating that \textit{``The KOC has to be dedicated to one particular area.''} If a creator covers all kinds of content in their account, for example, spanning from skincare to baby care, food, and vacation, in Celia's words, \textit{``such messiness in content themes makes potential consumers feel that they are not professional and less persuasive.''} This emphasis on specialized content aligns with creators like Lexie and Jasper who produce niche content. However, Celia noted that to be involved in their campaigns, \textit{``KOCs should not sound too expert.''} 
After all, KOCs are expected to blend in with the general RED user base in that their mundane content resonates with other users, even though they perform particular personas and produce advertisement-oriented content. With such precise requirements, marketers meticulously review every word in a KOC's posts. For content creators, these expectations are often opaque; they have to make sense of what marketers expect through feedback and revisions from marketers. Once their content is approved, content creators can then gain some monetary rewards, though these tend to be modest due to the large number of KOCs involved.

\subsubsection{Obscuring KOCs in Corporate Systems}\label{sec:corporation-measurement}

KOCs and their contributions are often rendered invisible, particularly in the clerical operations of corporate systems.

Before launching a campaign, marketers need to go through the procurement procedure to secure a budget that covers payments to KOCs, KOLs, celebrities, and the RED platform itself. Marketers are required to submit itemized budget plans to the procurement department, which then invites various social agencies to bid for service contracts, and the lowest bidders typically win. Usually, in addition to the itemized budget plans, marketers have to provide several sets of data that promise corporations a great return on their investment. To measure the return that content creators bring in marketing campaigns, corporations have developed several metrics, including ``cost per real engagement'' (CPRE) \footnote{``Real engagement'' means effective engagement. It means that the followers interact with the content creators by commenting, or staying on the page longer than a certain number of seconds} and ``cost per engagement'' (CPE). 
CPRE is calculated by the ratio of ``likes'' to followers of a specific content creator. It measures the ``stickiness'' of their followers, i.e., their ability to maintain follower engagement over time \cite{stickiness}\footnote{The ``stickiness'' concept is particularly popular in the technology industry when assessing a digital product.}.
CPRE indicates a long-term return on the investment of the content creators, whereas CPE assesses the immediate investment return of a creator by comparing the cost of a post to the number of ``likes'' the post receives. 
In the long run, the corporation evaluates its influencer marketing strategies on RED and decides its year-round budget plans largely based on these metrics.

Itemized budget plans and metrics help stabilize and standardize the marketers' practices and influencer marketing processes within corporations. Ironically, despite KOCs' essential role in producing advertorial content in campaigns, KOCs are often left out of the standardized processes purposefully. To avoid complicated clerical work, in actual practice, funds allocated for KOCs are merged with those for KOLs and celebrities. This is because, as Celia explained, \textit{``Because KOCs cost only little money.''} For marketers, the administrative effort required to manage KOCs does not justify their relatively minor financial impact on the budget items. In this way, the presence and contributions of KOCs are not documented in the corporation’s clerical system, instead existing in a grey area in marketing campaigns.  

What's more, the monetary rewards KOCs gain are not the exact amount marketers put in their budget plans; social agencies and MCN agencies usually take a cut, about 55\% of KOCs' rewards, according to Fan, a content creator focusing on the cosmetic area. This cut in KOCs' rewards is not documented. Despite their marginalization in corporate financial planning, involving KOCs in RED campaigns remains a common strategy in the advertising industry. KOCs' invisibility underscores the precarious nature of their roles, as their compensation and recognition depend on informal agreements within the triangular relationship among KOCs, the corporation, and the RED platform. In the following section, we turn to the RED platform and show how the platform further contributes to the invisibility of KOCs within these stabilized working processes.

\subsubsection{Excluding KOCs from the RED Platform's Governance}\label{sec:red-governance}

While encouraging users to share their everyday life experiences, RED maintains a cautious stance on advertisements. It closely monitors the balance between the two and has established clerical procedures and technological mechanisms to oversee content containing advertisements. In influencer marketing campaigns, marketers are required to submit detailed reports of their itemized advertorial content via RED’s embedded ``Deadilon'' (蒲公英\textit{pu gong ying}) platform. This ensures that RED can oversee which content creator is producing what type of advertorial content and when. At the same time, the platform makes sure it makes a profit on the specific advertorial content. If advertorial content is not reported, it may be subject to shadowbanning~\cite{duffy2023platform} or bans by RED's algorithm. Celia recalled instances where both Chinese and international brands were removed from the platform's search engine for ten days as punishment for unapproved advertisements.

KOCs are not subject to the same stringent surveillance, and this grants them more flexibility in creating advertorial content in collaboration with brands. Such kind of freedom provides marketers the convenience to develop KOC-oriented campaigns. Meanwhile, the platform only recognized KOCs' content as mundane and daily, rather than commercials. It indicates that KOC advertorials have little or no commercial value. Further, their efforts put into seeking advertorial opportunities and creating advertorial content are dismissed by the platform. Ironically, in addition to the reductions by social and MCN agencies, the platform takes another cut---as Fan revealed, about 10\% of the rewards---from marketers' payment to the KOCs. In this way, RED acts as a gatekeeper that takes ownership of KOCs' content and commercial potential.  
Hence, KOCs only gain 35\% of the commission fee the marketers plan to pay. The RED platform reinforces KOCs' invisibility in the standardized influencer marketing processes, putting KOCs in an underestimated position and devaluing their advertorial content.

\subsection{(Dis)empowerment: KOCs Stuck Among Platform Curation, Monetary Rewards, and Personal Pursuits}\label{sec:(dis)empowerment}

For content creators, being a KOC paradoxically limits their creative potential while empowering their personal pursuits. We demonstrate the dual nature of the KOC role by highlighting the challenges these creators encounter and how they respond to them as well as the personal pursuits they associate with this role.


\subsubsection{The KOC Role Disempowers Content Creation}

Performing the platform-curated persona and maintaining the KOC role can be exhausting and stressful. For both Jasper and Shirley, the uncertainty of advertorial collaborations and the ad-hoc nature of marketing campaigns added to their stress. They expressed concern that a shift in content or a drop in follower engagement could jeopardize their ability to monetize their platforms.

After six months as a ``good-looking'' content creator, Jasper grew tired of taking good-looking pictures daily, echoing Lexie's waning interest in aesthetic content on Instagram. He stated, \textit{``I don't know how far I can go taking daily photos, [...] but I think the platform algorithm has labeled me as an aesthetic content creator -- it might be hard to change to other areas.''} Jasper was unsure of how the algorithm would curate his content nor how to keep his audience engaged if he were to produce a different type of content.

Shirley also faced challenges in continually performing her local guide persona, so she experimented with new types of content. She said,
\textit{``The number of restaurants in Dallas is limited, I cannot only do restaurant review videos,''} When Shirley explored alternative topics for her RED content, she noticed that her followers responded positively with even more likes to her unboxing and product review videos.
However, logistical issues arose because she spent much of her time in Dallas while most of her followers were Chinese who lived in Mainland China. It was difficult for her to source products popular in China for her review videos in the US, and she had to figure out mailing services to ship products from China to the US, which affected the frequency of her product review videos. This challenge prompted Shirley to consider further diversifying her content further. Yet, she also worried that a ``messy'' account feed would make it difficult for her to maintain a clear focus or ``be vertical'' in any one area, potentially impacting her ability to keep her followers engaged and the likelihood of monetization.

\subsubsection{The KOC Role Empowers Personal Pursuits}

The KOC role empowers content creators to fight for monetization and economic opportunities, which are oftentimes integral to their personal pursuits in post-pandemic economic precarity. 

For example, Lexie reached out to the marketing team of an English-learning app for a promotional opportunity. She presented herself as a KOC in her communication with the company and convinced them that her followers were the app's target audience the app sought. Using RED's embedded analytical data about followers' demographics and the number of ``likes'' and ``comments,'' Lexie successfully negotiated a fair rate for her service as a KOC. The marketing team of the app was impressed and involved Lexie in their campaign.  Similarly, Jasper launched his first collaboration, a promotion for a fragrance brand, by showcasing the high quality and influence of his content that resonated well with his followers. While Lexie's and Jasper's stories seem like examples of straightforward success, such smooth experiences are predicated on their insider knowledge about the role of KOCs and how to position themselves as competent ones, which they gained from their prior social media industry working experience. Without such insights, content creators are in a disadvantageous position in negotiation with brands (or MCN agencies on behalf of brands) for brand promotion opportunities that bring monetary rewards. The seemingly accessible opportunities for personal flourishing through monetizing content are in fact inaccessible to many, if not most, content creators, as seen in Iris's experience. Iris used to work in the finance industry in China and then pursued her PhD degree in the US. Iris first gained thousands of likes with the persona of a math PhD student in math. Although Iris repeatedly experimented with different posts on math knowledge helpful for financial investment, the content seemed hard to resonate with RED platform users and she rarely received more than a two-digit number of ``likes.'' After several failed attempts to secure partnerships, Iris gave up seeking advertorial opportunities and abandoned her ambitions to become a KOC.

Although the empowerment content creators find through the KOC role is conditional on personal knowledge and experience, such empowerment supports them in facing unemployment and other societal pressures in the post-pandemic anxiety and predicament. Sylvia, a young woman who recently graduated with a Master's degree in industrial design and living in Xi'an, China, shared how being a KOC on RED helped her assert her independence when she introduced what she did for a living to her parents: \textit{``Becoming a KOC and making money from the platform helped me to negotiate with my parents. I can tell them that I can make money without a nine-to-five job.''} Similarly, Fei, a female music major college graduate in Dali, China chose to create music instrument instructional videos on RED instead of following the parent-expected path of becoming a teacher. Although she had not monetized her content, she was optimistic about her potential success on RED. She noted, \textit{``This [KOC] option holds my back, so that I do not fall behind my peers too much.''} Fei believed that she would find some relief once she could make money from RED.

Side hustles from the KOC role are also attractive to those who are already employed. Lexie and Jasper, who held full-time positions at multinational companies in Shanghai, were concerned about their prestigious yet modestly-paying jobs. During the period of economic slowdowns resulting from the COVID-19 pandemic, the possibility of being laid off has worried many young Chinese white-collar workers. It became necessary and even indispensable for Lexie and Jasper to pursue side gigs in order to maintain their middle-class lifestyles that were fueled by constant commercial consumption \cite{rofel2007desiring}. 
As we collected data for this project, some of our informants told us that more and more young Chinese tried to join the content creation industry and aspired to monetize by becoming a KOC on social media platforms. Despite being disadvantaged in the content creation labor market, the title of “KOC” has remained a powerful phrase and a promise in the daily lives of content creators. In a very specific way, this role empowers young Chinese to explore the possibilities of working and living in a highly precarious economy.



}
\major{\section{Discussion}

Our research documents and examines the labor processes involved in content creators becoming and performing as KOCs on RED. We also extend our analysis beyond the act of content creation to explore the digital ecosystem that fosters the KOC role, including marketers, corporations, and the platform itself. This analysis highlights how the digital cultural production industry often devalues and underestimates the contributions of content creators.

In this section, we introduce the concept of ``informal labor'' and articulate why KOC labor is, to some extent, necessarily informal. On the one hand, content creators purposefully select part of their daily lives and integrate it into their content creation, which sustains the personas curated by the platform and maintains their visibility. Such labor to blur the line between themselves and personas is invisible. On the other hand, while it is a seemingly accessible side hustle that is essential to marketing campaigns and the platform’s content culture, it often goes unrecognized. Paradoxically, despite these disadvantages, such informal labor is perceived as desirable and is even promoted by the state as a viable alternative in contexts of high unemployment and flexible employment policies. Our exploration not only sheds light on the conditions faced by KOCs and Chinese young people but also on the implications of this nuanced form of digital work in a platform-mediated society.

\subsection{Conceptualizing Informal Labor Showcased by KOCs}

\subsubsection{Marketers and Corporations Informalize KOC Labor}

We have demonstrated that content creators, on their path to becoming KOCs, perform platform-curated personas and produce niche content. They seem to master the tactics that appear to be effective and help them succeed in monetizing by using various tactics. Whether effective or not, such tactics were not based on intuition; rather, KOCs devoted much labor to researching, observing, experimenting, and leveraging other knowledge sets. For instance, presenting the platform-curated personas not only requires them to perform in front of the camera while internalizing the ideas and emotions of their personas. Erving Goffman's analytical lens and metaphor of front/back stages \cite{goffman2023presentation} has widely informed scholarship focusing on social media online impression management in CSCW/HCI and adjacent areas like communication and psychology. For these scholars, the backstage self is authentic, while the frontstage personas can be manipulated to convey certain impressions to the desired audiences~\cite{ellison2006managing, ward2017you, pitcan2018performing, haimson2021online, reddy2024teaspoon}. \hy{However, Goffman's analytical lens does not fully explain the relations of the true self and the performed persona in the case of KOCs.} For the content creators who aspire to become KOCs on RED, the personas they perform on the front stage and their true selves are not necessarily clearly diverged. For instance, Lexie is a professional in the marketing industry, and her online workplace persona stems from her true identity as an employee of a multinational corporation in Shanghai. Shirley introduced and reviewed restaurants in Dallas, and her personas spoke what the real Shirley was thinking: she recommended restaurants based on her true self’s tastes. The personas curated by the platform are one of many that KOCs live through in their daily lives. The longer content creators think, feel, and speak through these personas, the more their true selves and performed personas become interconnected. In other words, their true selves are put onto their RED personas, so that the latter can think and speak in a way that contributes to the content creation. KOCs are not full-time influencers who live by creating advertorial content; instead, they are sometimes involved in influencer marketing but always remain everyday users. Performative labor~\cite{wang2020chinese, woodcock2019affective, zhang2019live} is integral and intrinsic to their content creation labor, \hy{and such performances blur the line between their true selves and personas constantly.} 

In addition to the labor KOCs devote to performing their personas (with the goal of monetization), they have to conduct other kinds of labor to solidify and amplify their content; they must learn to use the platform’s tools, analyze what their followers are particularly interested in, and stay vertical to their needs~\cite{lobato2016cultural}. They actively make use of their knowledge gained from their work experiences, just like Lexie and Jasper used insider knowledge of the marketing industry. Content creators are responsible for testing and learning what kinds of content work better than others. Shirley, for example, experimented with several topics and spent months trying to find an engaging one to add to her local restaurant review content. Besides, KOCs have to figure out other trivial but critical details that impact the frequency of their content creation, like Shirley managing to ship products from China to the US. Importantly, the hint of creativity and novelty is always essential to creators’ content, which prompts them to keep learning how to inspire their followers. These kinds of labor are often invisible and unknown to their followers and other outsiders.

Given the labor that KOCs devote to monetizing, they are excluded from the standardized and stabilized influencer marketing campaigns that are predicated on metrics of content effectiveness (a standard procedure in the influencer industry \cite{bishop2021influencer}). Marketers and corporations rely on standardized procedures to guarantee their workflow efficiency and returns on investment. By contrast, outside of the clerical work and corporate systems, there is an undefined and unrecognized area, to which KOCs contribute substantially to support the smooth operation of marketing campaigns. This does not mean KOCs are outside of the purview of marketers; rather, marketers have changing yet specific expectations for them. For instance, we have shown that KOCs have to be attractive, persuasive, and have some level of, but not too much, expertise, so that they appear to be everyday platform users.

The labor put into social media content creation is tedious, economically uncompensated, and emotionally extracted~\cite{caplan2020tiered, o2019weapons, duffy2017not, duffy_having_2015}. What is more, such labor is often ``hidden'' \cite{duffy_visibility_2022, weidhaas2017invisible}, ``often-stigmatized,''~\cite{nakamura2015afterword} and even ``in vain''~\cite{abidin2016aren}. For Duffy and Sawey~\cite{duffy_visibility_2022}, the social media workers' labor is dismissed by external actors like their employers, or concealed by workers themselves to protect their identities \cite{duffy_visibility_2022}. In our case, KOCs’ labor is both informalized and rendered invisible by marketers, corporate systems, and digital platform governance. Building on Duffy and Sawey's theorization of invisible/hidden labor, we argue that the labor of KOCs is not only concealed but informal. \textbf{Informal labor is marked by its instability and intermittency, subject to challenges from any party at any time and vulnerable to the whims of others.} Crucially, this type of labor occurs in a nebulous zone, unrecognized and undocumented by legal contracts, paperwork, or digital records that might clarify labor relations; as a result, KOCs' precarity is both pervasive and persistent. We propose the concept of ``informal labor'' to offer a new lens to understand content creation labor that is indispensable yet unrecognized by the social media industry. \hy{We argue that the paradox constructed in the influencer marketing industry continuously reinforces KOCs' devalued informal labor and subordinated position.}  Borrowing the marketer Celia’s words, this kind of informal labor functions like ``tap water:'' self-maintaining and self-sustaining. It continuously feeds content into marketing campaigns and, by extension, the digital platform itself.

\subsubsection{The RED Platform Informalizes KOC Labor}

We highlight informal labor by investigating the role of the platform in the making of KOCs. For a platform that aims to create a digital community with which to share life experiences, RED closely surveils the ratio of content shared by everyday platform users to content identified as having embedded advertorials.

We have shown that RED has developed the management tool Dandelion, to keep records of the advertorial content that creators produce. However, it is noteworthy that KOCs involved in marketing campaigns are not reported through this management platform. However, this does not mean that RED does not extract economic benefits from KOCs. Platforms that mediate advertorial content and facilitate influencer marketing campaigns play the role of gatekeeper and seek rent from KOCs and marketers. Existing scholarly work has examined and criticized how platforms (e.g., YouTube) decide to share revenue with content creators~\cite{caplan2020tiered, kopf2020rewarding, burgess2018youtube}. Although platforms often encourage content creation, advocate for a participatory culture, and claim to play the role of ``partners,''  they judge, control, and limit content creators' monetization-seeking behaviors \cite{caplan2020tiered, kopf2020rewarding, burgess2018youtube}. Lobato~\cite{lobato2016cultural} calls on researchers to ``scrutinize specific logic of commercialization at work'' and decipher the ``monolithic operation and its effects'' (p.349). Our research on RED in the Chinese context illuminated the complicated relationship between content creators and platforms during the monetization process. KOCs are part-time everyday users and part-time advertorial creators, and their economic benefits are highly dependent on their relations with marketing campaigns. Paradoxically, platforms exclude KOCs from their technological design that surveil advertorial content but include them in their business models that profit from this design. Meanwhile, KOCs are controlled by the platform’s opaque algorithms, facing potential shadowbans~\cite{duffy2023platform, are2022shadowban} and other penalties. \hy{The platform also distills its cultural preferences by closely monitoring and curating the content and making sure it appears to be mundane and daily.} These practices that informalize and undervalue KOC labor further reinforce the platform’s power over KOCs, making a fake promise to content creators that they can easily monetize if they create some content.

Contextualizing KOCs within the platform economy, we argue that their creative work is one form of gig work, but with key nuances. Gig work hints at the nature of piecemeal and task-based, temporal, labor-intensive, and on-demand jobs, like food delivery, rideshare driving, and online pieceworking~\cite{kellogg2020algorithms, vallas_what_2020, gray2019ghost, ticona2018beyond, irani2015difference}. While the work conducted by KOCs is also piecemeal, task-based, temporal, and laborious, it is not fully on-demand. Rather, KOCs’ work is self-motivated and self-maintained, based on the ``do what you love'' mantra \cite{duffy_romance_2016}. Content creators aspiring to become KOCs often perform the necessary work of creating potentially monetizable content, for the role particularly long before they engage in marketing campaigns. Their ongoing labor devoted to content creation prior to campaigns sets them apart from typical ``on-demand'' gig work. For instance, Lexie’s sharing about her workplace experience provides possibilities to illustrate office supply products. Similarly, Shirley's restaurant review videos inherently promote some restaurants. The lack of formal labor agreements or contracts to acknowledge KOCs' continuous and ever-lasting labor \hy{further reinforces its informality.} Importantly, the platforms purposefully conceal this kind of informal labor to foster a casual, mundane, relatable platform culture.


\subsection{``Back to Labor'': From Informal Labor to Flexible Employment}

\subsubsection{A Renewed Focus of ``Back to Labor''}

The CSCW community has been conversing on the topic of ``back to labor'' which focuses on the labor process and coordination efforts within the labor processes~\cite{greenbaum_back_1996, tang_back_2023}. In conversation with this focus and its unique sociotechnical orientation, with our empirical case from China, we uncover how content creators navigate labor relations in the influencer marketing industry in order to monetize content.

By focusing on KOCs' labor processes, we have shown that their labor is informalized and invisiblized through interactions and relations with marketers/marketing campaigns, corporations investing in campaigns, and the RED platform.
Social media platforms make it seem easy to create content for profit. As we have demonstrated, RED promotes the idea that ``ordinary life’s charm touches people’s heart'' and encourages users to share their everyday experiences and engage others by posting pictures with text. 
RED also provides various online courses and tools to help users create popular content and monetize it.
On both discursive and technological levels, RED simplifies the process of content creation and monetization, promising creators an easy path to success.
However, the seemingly easy path to monetization is full of challenges. Our findings show that the challenges creators face stem not only from the need for creators to manage their visibility and follower base on impenetrable algorithm-dominated platforms, but also from navigating the unevenly distributed power dynamics among different stakeholders. The power relations triangulate among KOCs, the platform, marketers, and corporations that invest in campaigns, often to the detriment of KOCs. Marketers and corporations employ standardized management tools and paperwork to streamline and control the workflow, which often run against and exclude KOCs.

We have also demonstrated that content creators have to confront multiple layers of precarity, including unstable labor relations, downsized monetary rewards, and psychological impairment on their path to becoming KOCs. The platform’s easy access and well-designed technological tools simplify content creation and also lead creators to establish unrealistic expectations about easy monetization. As such, more and more people have been motivated to pursue content creation, which exacerbated the ongoing precarity. For this reason, while the seemingly lucrative role of the KOC can be empowering, it remains fraught with challenges. We caution potential KOCs about the challenges they may meet and advocate for greater transparency in the platform-mediated influencer marketing industry.

\subsubsection{Revisiting ``Flexible Employment'' in China}

We contextualize our case against the broad backdrop in which informal labor takes place, that is, post-pandemic economic precarity and national anxiety accompanied by the Chinese state’s promotion of ``flexible employment.''
Flexible employment (灵活就业\textit{ling huo jiu ye}) is not entirely new to China. Prior to the advent of the platform economy, between the 1990s and the 2000s, the Chinese government launched the policy of flexible employment so as to pacify the laid-off state workers and guide attention to market-based employment opportunities~\cite{shi2011, yang_unknotting_2015}. Decades later, during and after the COVID-19 pandemic, the discourse and policies on flexible employment have reemerged. This has been a critical juncture for many young Chinese who are experiencing unemployment, economic hardship, and anxiety. At the same time, more and more people have started to work from home, try different side hustles, and adjust to unconventional work arrangements, such as task-based temporary employment. Multiple platforms and the rising platform economy have shaped young Chinese's career paths (e.g., \cite{zhou2024trapped}). Mapping KOCs into the flexible employment discourse in China, we argue that the KOC role promises agency and fuels personal pursuit in times of precarity, yet it reinforces the requirement that each citizen be self-sustained and self-maintained, which actually exacerbates the precarious situation.

Interestingly, none of our content creator informants mentioned that the state’s flexible employment discourse or policies had influenced their work choices. Yet, in reality, flexible employment, as a discourse, is hegemonic, as it nudges and shapes citizens’ perceptions about what kinds of work they should pursue for their personal growth. For instance, content creator Sylvia, a new college graduate struggling with a competitive job market and her family’s expectations, found that becoming a KOC and earning income through content creation on RED significantly eased her financial and psychological burdens. 
Similarly, in the cases of Lexie and Jasper who both held full-time jobs, found that earning additional income through content creation as a side hustle helped them make ends meet.
They spontaneously took the task-based and temporal work mediated by the platform as a legitimate and effective way to seek economic benefit. This contrasts sharply with the previous generation's fixation on secure ``iron rice bowl'' jobs, and platform-mediated work reshapes young Chinese’s understanding of employment. Moreover, the benefits of flexible employment further support content creators to secure a decent social position in an uncertain economy.

Although the Chinese state promotes the flexible employment discourse to alleviate unemployment anxiety, it ultimately serves the state's own technopolitical goal of maintaining stability~\cite{chen2023maintainers, lindtner2020prototype}---a theme consistent with the reemployment project of the early 2000s~\cite{yang_unknotting_2015}. 
This governance strategy, which demands self-reliance from its citizens ~\cite{lindtner2020prototype, yang_unknotting_2015, wallis2013technomobility}, decades later, is now intersecting with technological advancements and the ``Internet Plus'' economy. It positions social media platforms as central to the state’s technopolitical agenda and its associated projects of modernization and prosperity,
but leaves citizens to manage the resulting precarity on their own. 

Our work illustrates the emerging roles of KOCs on social media platforms and highlights the informal labor processes that contribute to their precarious working conditions. Building on the aspirational view often associated with digital content creation \cite{duffy2017not}, which is usually unpaid, our findings reveal that corporations and platforms informalize content creation work and turn it invisible even when it is part of paid, standardized workflow and practices. By echoing Greenbaum’s~\cite{greenbaum_back_1996} call to the CSCW community, we underscore the urgency of revisiting labor processes, particularly at a time when digital platforms are becoming increasingly central to employment strategies. Hochschild's work on the emotional labor of flight attendants \cite{hochschild2019managed} has raised public awareness
and further informed feminist scholarship to fight for the recognition of different forms of devalued labor \cite{mcrobbie2011reflections, gill2002cool, suchman1995making}. This awareness has gradually influenced policymakers to address previously unpaid labor~\cite{buvinic2018invisible, weyrauch2011sound}. We advocate for a contextualized and nuanced examination of how labor is valued and compensated and urge for better protections and working conditions for informal laborers like KOCs in the social media content creation industry. By highlighting the case of KOCs on RED in China, we aim to contribute to broader discussions on informal labor. Further, we encourage future work to uncover other forms of informal labor in different platform-mediated industries and remain committed to understanding and addressing the evolving labor relations and working conditions that are increasingly shaped by algorithms, platforms, and technologies.
}
\major{\section{Conclusion}

We have shown how Key Opinion Consumers (KOCs), as content creators, are co-produced by marketers, corporations, and the RED platform. KOCs are instrumental in shaping potential consumers' impressions of brands and products as an integral part of marketing tactics; meanwhile, KOCs' content amplifies the mundane and daily life content popular on the RED platform, echoing the platform's discourse---``ordinary life's charm touches people's heart.'' They navigate the dynamics in the triangulated relations with the marketers and the platform in order to secure economic opportunities for producing advertorial content, and yet, the labor involved in producing such content is deliberately obscured to make it appear as spontaneous, ordinary user posts for the sake of marketing campaigns. Additionally, the commercial value of their work is often underestimated and overshadowed in corporate paperwork, platform technological mechanisms, and business models, resulting in and reinforcing inadequate recognition and compensation of KOCs. 

By bringing to light the labor processes of KOCs, our work illustrates how corporations and platforms informalize KOCs' content creation work and turn it invisible, even when such work is part of paid, standardized workflow and practices. We propose the concept of ``informal labor'' to underscore the precarious and intermittent nature of KOCs' work resulting from the informalized relations between KOCs and corporations/platforms. 
This concept offers a new lens to understand content creation labor that is indispensable yet unrecognized by the social media industry.
Additionally, we highlight the potential challenges and risks faced by content creators in China as they navigate the changing employment landscape. We hope to spark further examination of various forms of informal labor integral to the platform economy.
}

\bibliographystyle{ACM-Reference-Format}

\end{CJK*}
\end{document}